# A Resonant Chain of Four Transiting, Sub-Neptune Planets

Sean M. Mills[1], Daniel C. Fabrycky[1], Cezary Migaszewski[2,3], Eric B. Ford[4,5,6], Erik Petigura[7,8], Howard Isaacson[7]

[1] Department of Astronomy and Astrophysics, The University of Chicago 5640 S. Ellis Ave, Chicago, IL 60637, USA

[2] Institute of Physics and CASA*, University of Szczecin, Wielkopolska 15, 70-451 Szczecin, Poland

[3] Torun Centre for Astronomy, Nicolaus Copernicus University, Gagarina 11, 87-100 Torun, Poland

[4] Center for Exoplanets and Habitable Worlds,

[5] Department of Astronomy and Astrophysics, and

[6] Center for Astrostatistics, The Pennsylvania State University, University Park, PA 16802, USA

[7] University of California at Berkeley, Berkeley, CA 94720, USA

[8] California Institute of Technology, Pasadena, CA, 91125, USA

**Surveys have revealed many multi-planet systems containing super-Earths and Neptunes in orbits of a few days to a few months[1]. There is debate whether in situ assembly[2] or inward migration is the dominant mechanism of the formation of such planetary systems. Simulations suggest that migration creates tightly packed systems with planets whose orbital periods may be expressed as ratios of small integers (resonances)[3–5], often in a many-planet series (chain)[6]. In the hundreds of multi-planet systems of sub-Neptunes, more planet pairs are observed near resonances than would generally be expected[7], but no individual system has hitherto been identified that requires migration to form. Proximity to a resonance enables the detection of planets perturbing each other[8]. Here we report transit timing variations of the four planets in the Kepler-223 system, model them as resonant-angle librations, and com-**

pute the long-term stability of the resonant chain. The architecture of Kepler-223 is too finely tuned to have been formed by scattering, and our numerical simulations demonstrate that its properties are natural outcomes of the migration hypothesis. Similar systems could be destabilized by any of several mechanisms[5,9–11], contributing to the observed orbital-period distribution, where many planets are not in resonances. Planetesimal interactions in particular are thought to be responsible for establishing the current orbits of the four giant planets in the Solar System by disrupting a theoretical initial resonant chain[12] similar to that observed in Kepler-223.

Kepler-223 is a known four-planet system[13] orbiting around a slightly evolved (about 6-Gyr-old), Sun-like star (see Methods, Extended Data Fig. **1**). The low observational signal-to-noise ratio initially caused an incorrect identification of the orbital periods of this system[13,14], and has hitherto precluded its detailed characterization. For the analysis of transit timing variation (TTV), we use long cadence (29.4-min integrations) data, collected over the full duration of NASA's Kepler Space Mission from March 2009 to May 2013. Over this window, the ratios of the orbital periods (P) of planets b, c, d and e (named in alphabetic order from the interior, beginning with b) average $P_c/P_b = 1.3336$, $P_d/P_c = 1.5015$ and $P_e/P_d = 1.3339$[15]. We expect a system with periods so close to resonance to exhibit TTVs due to planet–planet interactions[8] (see Methods).

To measure TTVs, we bin the data into 3-month segments based on Kepler's observing quarters, confirm that the orbital periods are near resonances, and demonstrate the time-variable nature of the transits (Fig. **1**, Extended Data Fig. **2**, Extended Data Table **1**, and Methods). Phase folding the data and removing the TTVs allows the noisy transits to be identified easily by eye (Fig. 2).

The behaviour of the resonant chain can be characterized by its Laplace angles: $\phi_1 \equiv -\lambda_b + 2\lambda_c - \lambda_d$, $\phi_2 \equiv \lambda_c - 3\lambda_d + 2\lambda_e$ (for mean longitudes $\lambda_i$ and planets $i = $ b, c, d, e) and, for the whole system of four planets, $\phi_3 \equiv 2\phi_2 - 3\phi_1 = 3\lambda_b - 4\lambda_c - 3\lambda_d + 4\lambda_e$. Systems that are in resonance possess such librating Laplace angles, which ensures that two planets have a close approach when the other planets are far away, reducing chaotic interactions. The existence of a single four-body Laplace angle demonstrates that all the planets have close dynamical contact (with various three- and two-body resonances also present). We infer variations in the Laplace angles directly from the measured TTVs (see Methods and Extended Data Fig. 3). If we assume nearly circular orbits, the four years of TTVs in the data have recorded both angles performing nearly a full oscillation; $\phi_1$ librates between approximately $173°$ and $190°$ and $\phi_2$ librates between approximately $47°$ and $75°$.

To improve the treatment of the TTV signal and directly connect it to planetary dynamics, we integrate the N-body equations of motion for the four-planet system and explicitly model the photometric transit signals over the Kepler observing window (photodynamical modelling)[16]. We determine best estimates and uncertainties for the system parameters by performing five-body integrations of initial conditions from the resulting posterior distribution for more than $10^7$ orbits of the planets and retaining only parameter sets that remain stable (see Methods for details). We find that the planets all have masses of $3M_{\mathrm{Earth}}$–$9M_{\mathrm{Earth}}$ and radii of $2.5R_{\mathrm{Earth}}$–$5.5R_{\mathrm{Earth}}$ ($M_{\mathrm{Earth}}$ and $R_{\mathrm{Earth}}$ are the mass and radius of Earth, respectively; see Table 1). On the basis of these values and internal structure models[17], we determine that the composition of the planets varies from about 1% to 5% H/He by mass for the innermost planet to more than about 10% by mass for the outermost planet; that is, they are all sub-Neptunes. The density of the planets decreases with orbital semi-major axis, consistent with scenarios involving atmospheric loss due to stellar irradiation or formation in regions of increasingly cooler temperatures[18]. The eccentricities of the planets are relatively low (about 0.01–0.1) in configurations that are stable for more than $10^7$ orbits of the system. To fit the data acceptably, the eccentricities need to be slightly larger than in other systems of sub-Neptunes such as Kepler-11, whose eccentricities are less than about 0.02[19]. Because the eccentricities may be excited and stabilized by the resonances, the system can remain stable even though it is compact. The eccentricity of a planet is only loosely negatively correlated with its mass (from the TTVs in the data), so small changes in the allowed eccentricity will have a small effect on the posterior mass estimate, and removing eccentricity constraints would make the planets only slightly less dense.

Periods in a ratio close to 3:4:6:8 are maintained in all the stable, data-fitting solutions. The range of the ratios of the osculating periods of the planets implied by the observed TTVs over the Kepler window is typical for a resonant system. This range is narrower than that for a long-lived (more than about $10^7$ orbits), but circulating (non-resonant), solution (Extended Data Figs. **4** and **5**), suggesting that the system is currently in a state of libration. This libration might be temporary, and periods of Laplace-angle circulation might have occurred previously or might occur in the future for this system. However, requiring short-term Laplace-angle libration substantially increases the likelihood that a parameter set that acceptably fits the data represents a long-lived system (see Methods). Because (i) the orbital parameters of Kepler-223 are consistent with it being in a resonant state, (ii) solutions that are stable for 100 Myr exist within the parameter posteriors, and (iii) resonance greatly helps a system this compact to remain stable, we conclude that the system is probably a true resonant chain.

Planetary migration in a disk has been extensively studied and often leads to resonant chains of planets[3–6]. To examine the plausibility of the specific resonant chain observed in Kepler-223, we use a previously developed model[20] to simulate the migration of four planets within a gas disk. We find that four planets starting well wide of resonance migrate inwards and converge to the 3:4:6:8 chain of periods that we observe with certain choices of simulation parameters (Fig. **3**). Thus the Kepler-223 system is a plausible outcome of disk migration, but the full set of disk migration parameters and initial conditions that would lead to this system remains an open question.

In a migration scenario, systems trapped in resonances for which the orbital semi-major axes are small (less than about 0.5 au) can potentially be used to constrain the rate of disk photoe-vaporation and the lifetimes of disks, because a gaseous disk must exist in the 0.02–0.2 au range long enough for planets of moderate mass to migrate. It also provides constraints on turbulence and magnetic fields in the disk[21], and the structure of the disk that causes the planets to stop migrating[22]. An alternative to gas-disk migration for trapping planets into resonances is migration via planetesimal scattering[23]. It is possible for planetesimal scattering to migrate two planets in a convergent manner, establishing a resonance. However, this convergent migration would excite the eccentricities of the planetesimal population, which would probably prevent additional planets from joining the resonance[24]. The presence of a large volatile (greater than about 10% H/He by mass)[17] layer on the outer planets also suggests that the planets formed in the presence of a gas-containing disk at cool temperatures, further suggesting large-scale migration[18].

Several other exoplanet systems have (GJ 876[25]), or are speculated to have (HR 8799[20]), resonant chains, but these are composed of planets that are substantially more massive and have much greater orbital distances; hence, these observations may not be relevant to the formation of systems of close-in sub-Neptunes. Several Kepler systems are probably in a true resonance (as opposed to near resonance; for example, the 6:5 system Kepler-50 and the 5:4:3 system Kepler-60[26]); however, owing to the large number of known multi-planet systems, even if the orbital-period ratios of planets are essentially random, consistent with in situ, giant-impact formation, we would expect to observe some systems whose period ratios were near enough to integer values that they entered true dynamical resonances. By contrast, the precise conditions for the four-planet resonant chain of Kepler-223 cannot be accounted for by random selection of period ratios[7], and the system is probably too fragile to have been assembled by giant impacts[27].

The dynamical fragility of Kepler-223 suggests that resonant chains were precursors to some of the more common, non-resonant systems and that planet–planet scattering post-formation is probably an important step in creating the observed period distribution[10]. A model of the for-mation of the Solar System that has parallels with observed exoplanets involves the four giant planets entering a series of resonances, reaching their current configuration only after destabilization hundreds of millions of years later[12]. Numerical simulations for Kepler-223 indicate that only a small mass of orbit-crossing planetesimals is needed to move Kepler-223 off resonance[28], but that it could escape this fate if intrinsic differences in protoplanetary disks resulted in the lack of such a planetesimal population. In fact, various mechanisms including disk dissipation[9], planet–planet scattering[10], tidal dissipation[5] and planetesimal scattering[11] could break migration-induced resonances in the majority of exoplanet systems. It has been suggested that some multi- resonant systems (for example, Kepler-80, which has planetary pairs near, but not in, two-body resonances) might have undergone resonant disruption as a result of tidal dissipation, which would explain most of the period ratios that are slightly greater than resonant values in Kepler data[29,30]. It is possible that the Kepler-223 resonance has survived as a result of its relatively more distant innermost planet. Overall, we suggest that substantial migration of planets, including epochs of resonance that are typically only temporary, rather than in situ formation, leads to the final, observed planetary orbits for many close-in sub-Neptune systems.

**Main Text References**

**Acknowledgements** We thank Andrew Howard and Geoff Marcy for their role in obtaining spectra and Eric Agol, Jack Lissauer, and Jacob Bean for helpful comments on the manuscript. This material is based upon work supported by NASA under Grant Nos. NNX14AB87G (D.C.F.), NNX12AF73G (E.B.F.) and NNX14AN76G (E.B.F.) issued through the Kepler Participating Scientist Program. E.B.F received support from NASA Exoplanet Research Program award NNX15AE21G. D.C.F received support from the Alfred P. Sloan Foundation. C.M. was supported by the Polish National Science Centre MAESTRO grant DEC-2012/06/A/ST9/00276. The Center for Exoplanets and Habitable Worlds is supported by the Pennsylvania State University, the Eberly College of Science, and the Pennsylvania Space Grant Consortium. Computer simulations were run using the following facilities: the cluster "HAL9000" of the Faculty of Mathematics and Physics at the University of Szczecin; the cluster "Reef" installed in the Poznan Supercomputer Centre PCSS (computational grant No. 195); and the "Midway" cluster at University of Chicago Research Computing Center. Much of the data presented in this paper were obtained from the Mikulski Archive for Space Telescopes (MAST). STScI is operated by the Association of Universities for Research in Astronomy, Inc., under NASA contract NAS5-26555. Support for MAST for non-HST data is provided by the NASA Office of Space Science via grant NNX13AC07G and by other grants and contracts. The United Kingdom Infrared Telescope (UKIRT) is supported by NASA and operated under an agreement among the University of Hawaii, the University of Arizona, and Lockheed Martin Advanced Technology Center; operations are enabled through the cooperation of the Joint Astronomy Centre of the Science and Technology Facilities Council of the U.K. When the data reported here were acquired, UKIRT was operated by the Joint Astronomy Centre on behalf of the Science and Technology Facilities Council of the U.K. This work makes use of observations from the Las Cumbres Observatory Global Telescope Network, the Kepler Community Follow-up Observing Program (CFOP), and NASA's Astrophysics Data System (ADS). We thank NASA Exoplanet Science Institute (NExScI) and the University of California Observatories at University of California-Santa Cruz for their administration of the Keck Observatory. We extend special thanks to those of Hawaiian ancestry on whose sacred mountain of Mauna Kea we are privileged to be guests.

**Author Contributions**    S.M.M. performed the photodynamic, stability, tidal dissipation, and spectral evolution analyses and lead the paper authorship. D.C.F. designed the study, performed TTV and Laplace angle libration analysis, and assisted writing the paper. C.M. performed the migration analysis, assisted in initial data fitting, and contributed to the paper. E.B.F. advised on the DEMCMC analysis and paper direction. E.P. and H.I. obtained and analyzed spectra. All authors read and edited the manuscript.

**Author Information**    Kepler data is publicly available at http://archive.stsci.edu/kepler/. Reprints and permissions information is available at www.nature.com/reprints. The authors are not aware of any competing financial interests. Correspondence and requests for materials should be addressed to S.M.M. (sean.martin.mills@gmail.com).

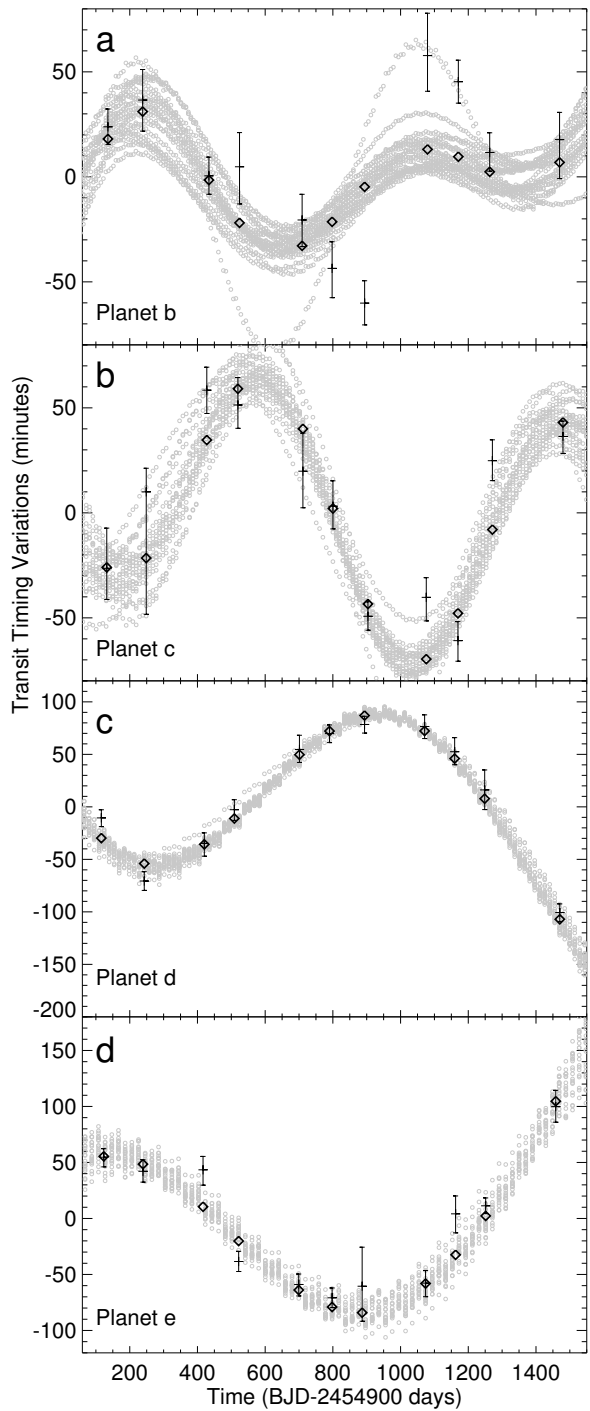

**Figure 1**: **Transit timing variations (TTVs) for all four planets with respect to a linear ephemeris. a-d**, Calculated transit times for planets b–e, respectively, come from a linear regression of the best-fit model transits. Open grey circles show the transit times from 20 different models that were stable over a $10^7$-year simulation. Black '+' symbols with $1\sigma$ error bars indicate the TTVs found by fitting quarterly binned data (see Extended Data Fig. **2**), and black diamonds are the corresponding points for the mean of the grey-circle models binned in the same manner. Where the noise causes large uncertainties, the photodynamic model may deviate from the binned data, but more accurately reflects the true TTVs. BJD, barycentric Julian date.

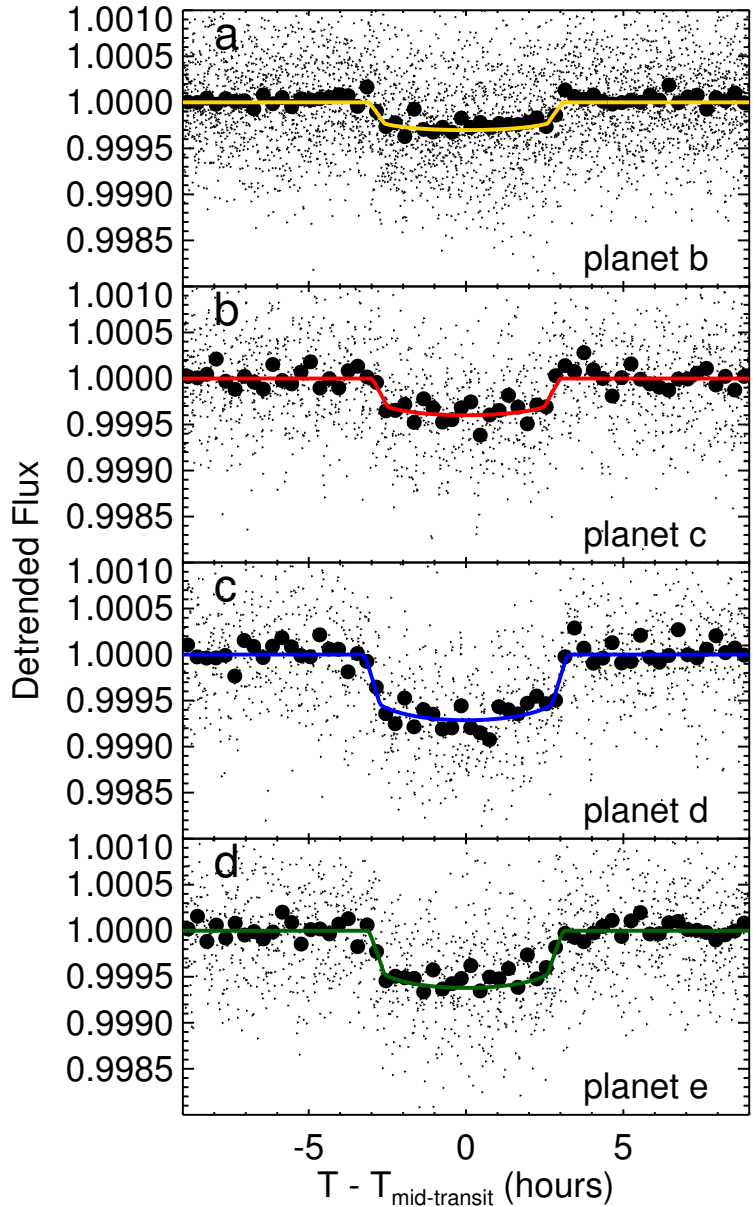

**Figure 2**: **Binned planet transits. a-d**, Photometry data near transits of planets b–e, respectively (small black triangles), binned together (large black circles) by phase-folding after removing the measured TTV for each quarter. Systematic trends have been removed and the flux normalized to 1.0 out of transit. The coloured lines are the best-fit transit models to the data.

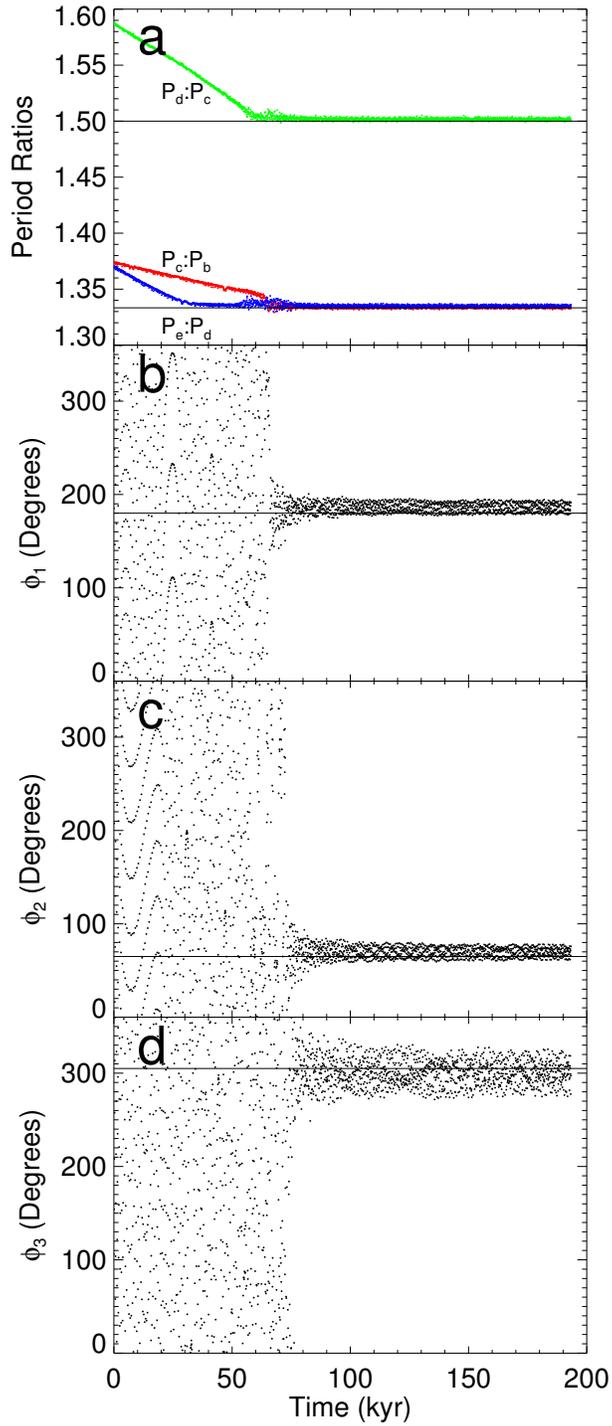

**Figure 3**: **A migration simulation which ends in a configuration matching the observed semi-major axis ratios and libration angle centers and amplitudes.** **a**, Time evolution of orbital-period ratios of planets b and c ($P_c/P_b$; red), c and d ($P_d/P_c$; green), and d and e ($P_e/P_d$; blue) in a migration simulation. **b–d**, Time evolution of the Laplace angles ($\phi_{1-3}$) defined in the text. The resonant angles and libration amplitudes that the planets end up in (indicated by the black horizontal lines) match those observed in the data (see, for example, Extended Data Fig. **3**).

| Parameter Name (Unit) | DEMCMC Result |
| --- | --- |
| Spectroscopic Stellar Mass ($M_\odot$) | $1.125^{+0.094}_{-0.073}$ |
| Stellar Radius ($R_\odot$) | $1.72^{+0.07}_{-0.14}$ |
| *Kepler-223 b Parameters:* | |
| $P$ (d) | $7.38449^{+0.00022}_{-0.00022}$ |
| $e$ | $0.078^{+0.015}_{-0.017}$ |
| $\lvert i - 90 \rvert$ (°) | $0.0^{+1.8}$ |
| $M$ ($M_{Earth}$) | $7.4^{+1.3}_{-1.1}$ |
| $R$ ($R_{Earth}$) | $2.99^{+0.18}_{-0.27}$ |
| $\rho$ (g/cm$^3$) | $1.54^{+0.63}_{-0.35}$ |
| *Kepler-223 c Parameters:* | |
| $P$ (d) | $9.84564^{+0.00052}_{-0.00051}$ |
| $e$ | $0.150^{+0.019}_{-0.051}$ |
| $\lvert i - 90 \rvert$ (°) | $0.0^{+1.3}$ |
| $M$ ($M_{Earth}$) | $5.1^{+1.7}_{-1.1}$ |
| $R$ ($R_{Earth}$) | $3.44^{+0.20}_{-0.30}$ |
| $\rho$ (g/cm$^3$) | $0.71^{+0.33}_{-0.20}$ |
| *Kepler-223 d Parameters:* | |
| $P$ (d) | $14.78869^{+0.00030}_{-0.00027}$ |
| $e$ | $0.037^{+0.018}_{-0.017}$ |
| $\lvert i - 90 \rvert$ (°) | $2.06^{+0.26}_{-0.32}$ |
| $M$ ($M_{Earth}$) | $8.0^{+1.5}_{-1.3}$ |
| $R$ ($R_{Earth}$) | $5.24^{+0.26}_{-0.45}$ |
| $\rho$ (g/cm$^3$) | $0.31^{+0.12}_{-0.07}$ |
| *Kepler-223 e Parameters:* | |
| $P$ (d) | $19.72567^{+0.00055}_{-0.00054}$ |
| $e$ | $0.051^{+0.019}_{-0.019}$ |
| $\lvert i - 90 \rvert$ (°) | $2.00^{+0.21}_{-0.27}$ |
| $M$ ($M_{Earth}$) | $4.8^{+1.4}_{-1.2}$ |
| $R$ ($R_{Earth}$) | $4.60^{+0.27}_{-0.41}$ |
| $\rho$ (g/cm$^3$) | $0.28^{+0.12}_{-0.08}$ |

**Table 1**: **Kepler-223 System Parameters.** Medians and 68% credible intervals for planet properties based on 2,008 $10^6$-year stable solutions with eccentricity priors as described in Methods: ($e_{b,max}, e_{c,max}, e_{d,max}, e_{e,max} = (0.212, 0.175, 0.212, 0.175)$ and fixed nodal angle $\Omega_j = 0$ for $j = b, c, d, e$ All values are valid at an epoch time $T_{\text{epoch}} = 800.0$ (BJD - 2,454,900). The stellar mass ($M_\star$) was held fixed in the differential-evolution Markov chain Monte Carlo (DEMCMC) simulation, but uncertainties in planetary mass were adjusted afterward to account for the quoted spectroscopic uncertainty in $M_\star$. $M_\odot$ and $R_\odot$ are the mass and radius of the Sun, respectively; $M_{\text{Earth}}$ and $R_{\text{Earth}}$ are the mass and radius of Earth. See Methods and Extended Data Table **2** for additional parameters and discussion.

**Methods**

**Stellar Properties**    To improve our knowledge of the Kepler-223 system, we obtained a spectrum of the host star on 10 April 2012 using the HIRES spectrometer[31] at the Keck-1 10 m telescope. These data are now publicly available at cfop.ipac.caltech.edu. After normalizing the continuum, we model the observed spectrum using synthetic spectra. Model spectra are generated by interpolating within a grid of synthetic spectra[32]. The resulting spectroscopic parameters for Kepler-223 are $T_{\rm eff} = 5821 \pm 123$ K, $\log(g) = 4.070 \pm 0.096$ dex, and [Fe/H] $= 0.060 \pm 0.047$ dex (where $g$ is the surface gravity in cgs units; and metallicities, [Fe/H] the logarithm of the ratio of iron to hydrogen in the star relative to that ratio in the sun).

To determine an age and mass of the star, we match the measured properties to $Y^2$ isochrones[33]. We ran a Markov chain Monte Carlo (MCMC) using the spectroscopic data and an interpolation of the $Y^2$ grid values as the model to obtain age $= 6.3^{+1.8}_{-1.7}$ Gyr and mass $= 1.125^{+0.095}_{-0.073} M_\odot$ (see ED Fig. **1**). Combining with $\log(g)$, we measure the stellar radius, $R_\star = 1.54^{+0.21}_{-0.18} R_\odot$ and stellar density, $\rho_\star = 0.31^{+0.12}_{-0.09} \rho_\odot$. We also derive a distance of $2.29^{+0.34}_{-0.34}$ kpc to Kepler-223 and find the mean flux on the planets to be $S_b = (492 \pm 47) S_0$, $S_c = (335 \pm 32) S_0$, $S_d = (195 \pm 19) S_0$, and $S_e = (133 \pm 13) S_0$, where $S_0 = 1377 W/m^2$ is the Earth's average insolation.

To determine the size of model-dependent uncertainties, we compare our results to an independently developed, publicly available method for computing $M_\star$, $R_\star$, and age using the Dartmouth isochrones[34] (available at: https://github.com/timothydmorton/isochrones). All 3 values are consistent within the 1-$\sigma$ error bars, so we conclude that our measurements are robust and model-dependent errors are small compared to our quoted uncertainties. We also use a stellar population synthesis model, TRILEGAL[35], with the default galaxy stellar distribution and population as described therein, to demonstrate the best fit mass and uncertainties described above are essentially unaffected by reasonable priors, so we keep flat priors in all stellar parameters.

**TTVs.** To measure TTVs, we begin by detrending the simple aperture photometry (SAP) flux data from the Kepler portal on the Mikulski Archive for Space Telescopes (MAST). For long-cadence data (quarters 1-8), we fit the amplitudes of the first five cotrending basis vectors (largest magnitude vectors from a Singular Value Decomposition of the photometry for a given CCD channel) to determine a baseline. We discard points marked as low quality (quality flag $\geq 16$). For short cadence data (58.8 second integrations, quarters 9-17), cotrending basis vectors are not available. Instead, we first masked out the expected transit times of a preliminary model, plus 20% of the full duration of each transit intending to account for possible additional timing variations; then we fit a cubic polynomial model with a 2-day width centered within half an hour of each data-point, to determine its baseline. In both cases, the baseline remains dominated by instrumental systematics which are time-variable; thus we divide the flux by this baseline.

In computing TTVs, we only use data for transits which do not overlap with another planetary transit, i.e., with two transit midtimes falling within 1 day of each other, according to a preliminary model (data with overlapping transits is modeled directly by the photodynamic method described later). To determine transit times, we first fit transit parameters (period, transit mid-time, planet-to-star radius ratio, transit duration, impact parameter, and limb darkening coefficient) to the entire long-cadence dataset. Second, we refit each quarter using the globally-determined values for all parameters except for transit mid-time, which is solved for. Thirdly, we refine the transit shape parameters and slide the refined transit model in time through the data for each planet in each quarter, computing the goodness-of-fit statistic $\chi^2$ in steps of $0.001$ days. The values of that numerical $\chi^2$ function which are within $1.0$ of the minimum are fit with a parabola, the minimum of which we adopt as our best estimate of the mid-time. The time shifts in each direction at which the $\chi^2$ function rises by $1$ and $9$ above the minimum are adopted as narrow and conservative error bars. If the likelihood surface of the mid-time parameter were Gaussian, these values would correspond to $1\sigma$ and $3\sigma$ estimates. ED Table **1** reports the average time of the transits that were combined to make each measurement, the best-estimate, and uncertainty estimates of these time shifts. Once phased at these transit times, the transit lightcurves are shown in Fig. **2**. These transit times are also represented graphically in ED Fig. **2** as the horizontal error bar. Planets c, d, and e all have fluctuations visible by eye over the data set. These data constitute our transit timing measurement, which does not depend on the photodynamical model we develop subsequently, nor are used therein.

We use these transit times to estimate the Laplace critical angles[36] and their evolution. To do so, we note that for circular orbits the mean longitude, $\lambda$, is a linear function of time, $t$, related to the transit period, $P$, and a specific mid-time, $T_0'$, as:

$$\lambda = 2\pi[1/4 + (t - T_0')/P].\tag{1}$$

In place of $T_0'$, we may use $T_0 + \Delta T_0$, where $P$ and $T_0$ define the linear ephemeris on which the quarterly $\Delta T_0$ of ED Table **1** are based. Then for Laplace's critical angles we have:

$$
\begin{aligned}
\phi_1 &= -\lambda_b + 2\lambda_c - \lambda_d, & (2) \\
&= 2\pi\{\frac{T'_{0b}}{P_b} + \frac{-2T'_{0c}}{P_c} + \frac{T'_{0d}}{P_d} + t(\frac{2}{P_c} - \frac{1}{P_b} - \frac{1}{P_d})\}, & (3) \\
&= 2\pi\{0.4750 + 2.39834 \times 10^{-5}(t - 2454900\,BJD) + \frac{\Delta T_{0b}}{P_b} - \frac{2\Delta T_{0c}}{P_c} + \frac{\Delta T_{0d}}{P_d}\}, & (4)
\end{aligned}
$$

and similarly,

$$
\begin{aligned}
\phi_2 &= \lambda_c - 3\lambda_d + 2\lambda_e, & (5) \\
&= 2\pi\{-\frac{T'_{0c}}{P_c} + \frac{3T'_{0d}}{P_d} - \frac{2T'_{0e}}{P_e} + t(\frac{1}{P_c} - \frac{3}{P_d} + \frac{2}{P_e})\}, & (6) \\
&= 2\pi\{0.1135 + 8.7366 \times 10^{-5}(t - 2454900\,BJD) - \frac{\Delta T_{0c}}{P_c} + \frac{3\Delta T_{0d}}{P_d} - \frac{2\Delta T_{0e}}{P_e}\}. & (7)
\end{aligned}
$$

These values are plotted in ED Fig **3**. The values are not constant, and the data appear to have sampled a minimum and maximum value of a libration cycle due to a restoring torque. The specific values are sensitive to phase shifts due to eccentricity-vector precession; the libration centers may be different by about $30°$ if the eccentricities are as high as $0.1$.

**Photodynamic Inputs** A Newtonian photodynamic model similar to existing models[37], but developed independently, was used for a dynamical analysis of this system. To find the most likely parameter values and uncertainties in the system, we run a differential evolution Markov Chain Monte Carlo (DEMCMC)[38] to compare model output for different system parameters to observed long and short cadence Kepler data, as well as spectroscopic data of the star. The TTV signal (Fig. **1**), which is here constrained by the photometry directly, detects the gravitational perturbations due to planet mass. Combined with transit shape information, this constrains eccentricities and provides significant ($\sim 10-30\%$) mass detections for all bodies.

Each planet has seven parameters: $\vec{p_i} = [P, T_0, e\cos(\omega), e\sin(\omega), i, \Omega, R_p/R_\star, M_p/M_\star]$ in which $P$ is the period, $T_0$ is the midtransit time, $e$ is eccentricity, $i$ is inclination, $\omega$ is the argument of periastron, $\Omega$ is the nodal angle, $R_p/R_\star$ is the planet-to-star radius ratio, and $M_p/M_\star$ is the planet-to-star mass ratio. The star has 5 parameters: $\vec{p_\star} = [M_\star, R_\star, c_1, c_2, dilute]$, where $c_i$ are the two quadratic limb-darkening coefficients and $dilute$ is the amount of dilution from other stars. Since photometry constrains only stellar density, and not mass and radius individually, we fix $M_\star$ at the best fit value found from spectroscopy and convolve the mass distribution with the DEMCMC posteriors when reporting final values.

We fix $\Omega = 0$ for all planets since the data do not sensitively measure mutual inclinations. We note that the typical mean mutual inclination (MMI) of Kepler systems, $\sim 1.8°$, implies near coplanarity[7]. Additionally, multiplanet systems with higher mutual inclinations between planetary orbital planes are correlated with instability[39], and we expect any observed system to be at least quasi-stable. Although for some pairs of planets photometry determines whether their inclinations are on the same side of $90°$[40,41], in preliminary runs we find no preference for either conclusion. Thus we exoplore only $i > 90°$ for each planet to reduce the volume of symmetrical parameter space. The value for the stellar limb darkening coefficient $c_2$ was chosen as 0.2 as this is near the median value for stars in the 4000K to 6500K range in the Kepler bandpass[42], and for low signal-to-noise ratio (SNR) transits like that in Kepler-223, a single limb darkening parameter is sufficient to match transit shape[43,44].

United Kingdom Infrared Telescope (UKIRT) archives reveal that there are two objects within 2" of the position specified by the Kepler Input Catalog (KIC)[45]. The brighter of the two objects has a distance < 0.2" from the KIC position and has a predicted Kepler magnitude of 15.4932, based on UKIRT archives' formula to convert their measured J band magnitudes to a Kepler magnitude[46]. This value is 0.1492 magnitudes fainter than that reported in the KIC (15.344). The second object is 1.937" away from the KIC location, but is about 8 times fainter. The sum of these two objects has a predicted intensity in the Kepler bandpass equal to 98.2% of the intensity of the object reported by the KIC. Faulkes Telescope North (FTN) imaging confirms the dual nature of the Kepler-223 object[47]. Speckle imaging done at WIYN observatory indicate no additional bodies between approximately 0.2" and 1.9" of the brighter object[48]. Since the fainter of the two objects contributes approximately 11.202% of the light in the Kepler bandpass, we perform our DEMCMC runs with the dilution fixed at 0.11202.

**Photodynamic Fits** Beginning the DEMCMC by distributing parameters over the entire 30-dimensional prior is computationally untenable for this problem as it would take an excessively long time for the parameter sets, $\{\vec{p}\}$, of the DEMCMC to escape local minima and reach the global minimum. Instead we begin the DEMCMC by taking a four-planet solution found by exploration using migration-assembly solutions, $\vec{p_0}$, which approximately matches the observed data, and forming a set of 48 30-parameter vectors, $\{\vec{p}\}_0$, by adding 30-dimensional Gaussian noise to $\vec{p_0}$. We allow each set to explore the parameter space, and in order to eliminate any effects of the choice of $\vec{p_0}$, we wait until the DEMCMC chains are converged and then remove a 'burn-in' period, that is, the portion that is dependent on the choice of $\{\vec{p}\}_0$.

In the DEMCMC, a given choice of planetary parameters is accepted or rejected based on the data over the Kepler observing window (about 4 years), and does not take into account the long-term evolution of a system with such parameters. It is not computationally tenable to numerically integrate each model for the age of the Kepler-223 system during the DEMCMC run. Thus the DEMCMC posterior includes solutions which acceptably fit the data, but which go unstable shortly after. To prevent our posterior parameter estimates from representing unstable solutions, we take two steps to encourage stability. First, we do not allow the DEMCMC to explore any solutions where the orbits of two adjacent planets cross (which generates a posterior we call $\mathcal{C}1$). This was implemented by allowing the DEMCMC to explore a limited range of eccentricities for each planet: $(e_{b,max}, e_{c,max}, e_{d,max}, e_{e,max}) = (0.212, 0.175, 0.212, 0.175)$, with the symmetry of values due to the resonant chain structure of the periods (posteriors can be found in ED Table **2** and best-fits in ED Table **3**). Retrospectively, this eccentricity prior is justified because mean eccentricities higher than $0.1$ are very rarely stable (ED Fig. **6**). Further, the similarity between the $10^6$ year eccentricity-stability distribution and the $10^7$ year distribution indicates that using either as a proxy for stable solutions will yield comparable results.

To assess the stability of the solutions in the posterior distribution, we selected 500 random draws from the $\mathcal{C}1$ posterior and numerically integrated each of these solutions for $10^7$ years, more than $10^8$ orbits of the outermost planet. We used the MERCURY symplectic integrator[49] and stopped integration if a close encounter between any two bodies occurred. 30% of systems lasted the entire $10^7$ year integration. We randomly selected 25 of the systems which lasted $10^7$ years and numerically integrated them for an additional $9 \times 10^7$ years, or until a close encounter, with 64% of them lasting $10^8$ years. The age of the Kepler-223 star is about $6 \times 10^9$ years. We expect the planets to have reached their current configuration by migration through a disk within only a few million years, corresponding to gas disk lifetimes[50], suggesting that the current planet configuration has also survived for about $6 \times 10^9$ years. However, integrating for this long is not computationally feasible for this study. We note, though, that other numerical stability studies[10] predict that systems are approximately equally likely to go unstable in bins of log[time], implying approximately 12% of the tested systems (and thus approximately 12% of the systems in the $\mathcal{C}1$ posterior) remain stable on Gyr timescales. This fraction is high compared to a modeled population of compact, sub-Neptune systems, which are destabilized by mean motion resonances (MMRs) on a shorter timescale[10]. However, in such simulations there are generally a few bodies not engaged in the resonance; here all four bodies are involved in the resonance, remaining stable despite MMRs exciting eccentricities. We also note that MERCURY is a Newtonian physics integrator, but adding a suitable General Relativistic (GR) potential term, $U_{GR} = -3((GM_\star)/(c\ r))^2$, where $c$ is the speed of light and $r$ is the distance from a planet to the star[51] does not change our long-term stability results based on 100 trials (32 stable, 68 unstable).

To develop a second posterior based on parameters more likely to be stable, we randomly drew 5,000 parameter sets found in the posterior of $\mathcal{C}1$, and numerically integrated each of these solutions for $10^6$ years (more than $10^7$ orbits of the outermost planet). This allows the problem to be computationally feasible, while still allowing for a great enough number of draws that we have sufficient statistics for parameter estimates. We retained only those parameter sets that remained stable at least this long (2,008 in total) to form a second posterior representative of physical, i.e. stable, solutions and call it $\mathcal{C}2$. Future discussions of parameters and the data table in the main text (Table **1** and Fig. **1**) use this posterior ($\mathcal{C}2$) as we judge it to be the optimal combination of selecting stable solutions which match the observed Kepler data, while avoiding discarding plausible parameter space due to further assumptions. The general shape of the eccentricity distribution remaining after $10^6$ years does not change markedly compared to solutions stable for an order of magnitude longer (see ED Fig. **6**) and is thus unlikely to change dramatically over the $\sim$6-Gyr age of the system. We note that the instability regions near the best-fit values discovered by our parameter fits suggest the ease with which the system, and others like it, could be moved out of resonance by only small perturbations such as evaporation of the protoplanetary disk or other means[9,28].

Kepler-223 appears to possess two librating Laplace angles between the inner three and outer three planets, as discussed earlier. Migration simulations suggest that very large Laplace angle libration amplitude is unlikely in stable solutions. Further, in stable solutions in the $\mathcal{C}2$ posterior, we note long-lived (up to $\sim 10^5$ year) Laplace angle libration is likely to occur. To get another estimate of the system's parameters while balancing computational efficiency and a stricter stability constraint, we ran a third DEMCMC. For this run, at every step in the DEMCMC we integrate the parameter initial conditions for 100 years ($\gtrsim 5$ secular oscillations) and penalize Laplace angle oscillation amplitudes that grew too large, in addition to fitting the data. We call the posterior from this run $\mathcal{C}3$. Our Laplace angle criteria in $\mathcal{C}3$ are designed to penalize large libration amplitudes, and the speed at which the amplitudes grow. If the total range in Laplace angles, $\Delta\phi_1$ or $\Delta\phi_2$, exceeds a cutoff value $K_1$ over the integration time ($T_{max}$ in years), then the time at which that occurred is recorded ($T_{runaway}$). A value $-1 + (T_{runaway}/T_{max})^{-2}$ is added to the $\chi^2$. All $\chi^2$ values were also penalized by a an additional term equal to $(\Delta\phi_i - V_i)^2$ for each $\Delta\phi_i > V_i$ or 0 if $\Delta\phi_i < V_i$ for specified angles $V_i$, $i = 1, 2$ in degrees and with $\Delta\phi = \phi_{max} - \phi_{min}$. This way if the Laplace angles were well enough behaved not to run away, but either or both still grew in amplitude above specified values for each angle, $V_1$ and $V_2$, then a $\chi^2$ penalty was assigned, and the parameters were less likely to be accepted. We impose no direct eccentricity constraint. We report $\mathcal{C}3$ with $(T_{max}, K_1, V_1, V_2) = (100\text{yr}, 170°, 30°, 50°)$, where the numbers are roughly based on the results of migration and DEMCMC results which had long-term libration (see ED Table **2**).

Running a similar stability check for $\mathcal{C}3$ as for $\mathcal{C}1$ by choosing 300 chains from the posterior distribution resulted in 100% of the parameter sets lasting $10^7$ years. Ten parameter sets were numerically integrated for $10^8$ years, and 100% of those also survive with no close encounters. These results indicate that this method is effective at finding stable solutions. Comparing this to the stability results for $\mathcal{C}1$ in which only 19% of solutions were stable for $10^8$ years as described above, our argument that the resonance does encourage stability is strengthened. Nevertheless, this method cannot be guaranteed to reject all unstable systems (because they might pass this test) or to include all stable ones (because some systems could remain stable for a very long time, but have large changes in Laplace angle); see ED Fig. **4**. This posterior has lower eccentricities, but because we assume short-term resonance for this fit, we do not take it as our nominal fit.

**Future Observations**  We predict future transit times and uncertainties by averaging the predicted transits from 152 $10^7$-year-stable solutions from the $\mathcal{C}1$ posterior. We report transit times quarterly for 10 years, including over the Kepler observing window, in Supplementary Information.

**Code Availability**  The code used for migration simulations is publicly available and can be downloaded from the online edition of this article.

**Methods References**

| $\bar{t}$ − 2454900 (BJD) | −3σ | −σ | Best | +σ | +3σ |
|---|---|---|---|---|---|
| *Kepler-223b: P* = 7.3840154 *days, T₀* − 2454900 *(BJD)* = 70.49489 | | | | | |

| $\bar{t}$ − 2454900 (BJD) | −3σ | −σ | Best | +σ | +3σ |
|---|---|---|---|---|---|
| *Kepler-223b: P* = 7.3840154 *days, $T_0$* − 2454900 *(BJD)* = 70.49489 | | | | | |
| 123.32662 | -0.0354 | -0.0058 | -0.0006 | 0.0059 | 0.0316 |
| 239.02516 | -0.0517 | -0.0103 | 0.0137 | 0.0101 | 0.0423 |
| 416.51724 | -0.0200 | -0.0061 | -0.0010 | 0.0062 | 0.0210 |
| 521.69775 | -0.0628 | -0.0123 | 0.0068 | 0.0113 | 0.0342 |
| 699.18988 | -0.0470 | -0.0088 | -0.0010 | 0.0084 | 0.0370 |
| 797.79657 | -0.0417 | -0.0097 | -0.0123 | 0.0088 | 0.0243 |
| 886.54260 | -0.0343 | -0.0072 | -0.0187 | 0.0074 | 0.0617 |
| 1073.89526 | -0.0500 | -0.0118 | 0.0730 | 0.0140 | 0.1150 |
| 1162.64136 | -0.0542 | -0.0071 | 0.0692 | 0.0071 | 0.0708 |
| 1251.38745 | -0.0217 | -0.0062 | 0.0507 | 0.0065 | 0.0333 |
| 1458.46155 | -0.0379 | -0.0129 | 0.0659 | 0.0090 | 0.0241 |
| *Kepler-223c: P* = 9.8487130 *days, $T_0$* − 2454900 *(BJD)* = 71.37624 | | | | | |
| 116.10564 | -0.0362 | -0.0103 | -0.0168 | 0.0133 | 0.0518 |
| 242.86336 | -0.0683 | -0.0405 | 0.0023 | 0.0077 | 0.0747 |
| 420.32413 | -0.0254 | -0.0077 | 0.0264 | 0.0076 | 0.0476 |
| 509.05453 | -0.0266 | -0.0077 | 0.0166 | 0.0090 | 0.0304 |
| 701.30371 | -0.0585 | -0.0121 | -0.0155 | 0.0126 | 0.0535 |
| 790.03418 | -0.0302 | -0.0075 | -0.0318 | 0.0084 | 0.0268 |
| 886.15869 | -0.0173 | -0.0046 | -0.0737 | 0.0048 | 0.0537 |
| 1071.01367 | -0.1404 | -0.0078 | -0.0766 | 0.0066 | 0.0226 |
| 1148.65283 | -0.0411 | -0.0067 | -0.0959 | 0.0064 | 0.0349 |
| 1252.17163 | -0.0392 | -0.0067 | -0.0418 | 0.0068 | 0.0548 |
| 1470.30054 | -0.0361 | -0.0056 | -0.0449 | 0.0051 | 0.0179 |
| *Kepler-223d: P* = 14.7883997 *days, $T_0$* − 2454900 *(BJD)* = 109.76775 | | | | | |
| 132.10997 | -0.0416 | -0.0058 | 0.0376 | 0.0054 | 0.0134 |
| 248.65308 | -0.0221 | -0.0062 | -0.0169 | 0.0063 | 0.0229 |
| 427.57138 | -0.0351 | -0.0084 | -0.0099 | 0.0070 | 0.0169 |
| 519.49268 | -0.0285 | -0.0070 | 0.0035 | 0.0066 | 0.0245 |
| 711.54260 | -0.0260 | -0.0086 | 0.0240 | 0.0094 | 0.0420 |
| 800.18097 | -0.0226 | -0.0060 | 0.0256 | 0.0057 | 0.0194 |
| 898.66815 | -0.0192 | -0.0057 | 0.0212 | 0.0055 | 0.0238 |
| 1077.35193 | -0.0530 | -0.0080 | 0.0020 | 0.0077 | 0.0210 |
| 1169.50781 | -0.0354 | -0.0085 | -0.0236 | 0.0093 | 0.0286 |
| 1271.27771 | -0.0272 | -0.0131 | -0.0578 | 0.0132 | 0.0328 |
| 1483.43542 | -0.0298 | -0.0061 | -0.1612 | 0.0057 | 0.0302 |
| *Kepler-223e: P* = 19.7213435 *days, $T_0$* − 2454900 *(BJD)* = 68.10686 | | | | | |
| 135.47421 | -0.0303 | -0.0060 | -0.0067 | 0.0053 | 0.0187 |
| 238.21753 | -0.0232 | -0.0067 | 0.0022 | 0.0072 | 0.0458 |
| 433.78842 | -0.0542 | -0.0095 | 0.0302 | 0.0084 | 0.0298 |
| 524.27625 | -0.0244 | -0.0061 | -0.0106 | 0.0063 | 0.0296 |
| 709.21222 | -0.0432 | -0.0071 | 0.0022 | 0.0065 | 0.0208 |
| 797.82037 | -0.0240 | -0.0060 | 0.0090 | 0.0061 | 0.0240 |
| 893.81256 | -0.0357 | -0.0216 | 0.0297 | 0.0242 | 0.0513 |
| 1079.88989 | -0.0662 | -0.0083 | 0.0602 | 0.0078 | 0.0308 |
| 1170.71301 | -0.1067 | -0.0118 | 0.1167 | 0.0110 | 0.0453 |
| 1263.01343 | -0.0252 | -0.0049 | 0.1352 | 0.0049 | 0.0188 |
| 1469.48169 | -0.0393 | -0.0097 | 0.2283 | 0.0100 | 0.0467 |

**Extended Data Table 1**: **Mean Kepler-223 quarterly TTVs.** Transit times and TTVs (in days) for each planet found by binning the data quarterly and iteratively solving for transit shape as described in the Methods. Mean transit time in the quarter is given in the first column followed by the measured TTV and uncertainties.

| Parameter Name (Unit) | Eccentricity Prior ($\mathcal{C}1$) | Eccentricity Prior and Stability ($\mathcal{C}2$) | Laplace Angle Constraint ($\mathcal{C}3$) |
|---|---|---|---|
| *Stellar Parameters:* | | | |
| $R_*(R_\odot)$ | $1.714^{+0.079}_{-0.165}$ | $1.72^{+0.07}_{-0.14}$ | $1.622^{+0.078}_{-0.070}$ |
| $M_*(M_\odot)$ | 1.125 (fixed) | 1.125 (fixed) | 1.125 (fixed) |
| $c_1$ | $0.54^{+0.11}_{-0.10}$ | $0.54^{+0.11}_{-0.09}$ | $0.57^{+0.11}_{-0.10}$ |
| $c_2$ | 0.2 (fixed) | 0.2 (fixed) | 0.2 (fixed) |
| dilution | 0.11202 (fixed) | 0.11202 (fixed) | 0.11202 (fixed) |
| *Kepler-223 b Parameters:* | | | |
| P (d) | $7.38454^{+0.00024}_{-0.00028}$ | $7.38449^{+0.00022}_{-0.00024}$ | $7.38453^{+0.00024}_{-0.00024}$ |
| $T_0$ (BJD-2454900) | $801.5145^{+0.0044}_{-0.0047}$ | $801.5155^{+0.0044}_{-0.0046}$ | $801.5133^{+0.0042}_{-0.0045}$ |
| $e \cdot \cos(\omega)$ | $0.057^{+0.034}_{-0.031}$ | $0.054^{+0.031}_{-0.022}$ | $0.035^{+0.014}_{-0.016}$ |
| $e \cdot \sin(\omega)$ | $0.052^{+0.026}_{-0.135}$ | $0.047^{+0.020}_{-0.039}$ | $-0.004^{+0.029}_{-0.034}$ |
| $|i - 90|$ (°) | $0.0^{+1.7}$ | $0.0^{+1.8}$ | $0.0^{+1.4}$ |
| $\Omega$ (°) | 0.0 (fixed) | 0.0 (fixed) | 0.0 (fixed) |
| $M/M_*$ | $0.0000196^{+0.0000034}_{-0.0000031}$ | $0.0000221^{+0.0000034}_{-0.0000031}$ | $0.0000201^{+0.0000027}_{-0.0000026}$ |
| $R/R_*$ | $0.01596^{+0.00055}_{-0.00053}$ | $0.01597^{+0.00055}_{-0.00053}$ | $0.01584^{+0.00053}_{-0.00053}$ |
| *Kepler-223 c Parameters:* | | | |
| P (d) | $9.84584^{+0.00085}_{-0.00053}$ | $9.84564^{+0.00052}_{-0.00051}$ | $9.84613^{+0.00046}_{-0.00051}$ |
| $T_0$ (BJD-2454900) | $800.1461^{+0.0049}_{-0.0040}$ | $800.1459^{+0.0050}_{-0.0039}$ | $800.1489^{+0.0061}_{-0.0047}$ |
| $e \cdot \cos(\omega)$ | $0.030^{+0.050}_{-0.047}$ | $0.029^{+0.041}_{-0.038}$ | $-0.010^{+0.019}_{-0.022}$ |
| $e \cdot \sin(\omega)$ | $0.134^{+0.027}_{-0.156}$ | $0.139^{+0.020}_{-0.050}$ | $0.060^{+0.033}_{-0.038}$ |
| $|i - 90|$ (°) | $0.0^{+1.4}$ | $0.0^{+1.3}$ | $0.0^{+1.5}$ |
| $\Omega$ (°) | 0.0 (fixed) | 0.0 (fixed) | 0.0 (fixed) |
| $M/M_*$ | $0.0000157^{+0.0000048}_{-0.0000038}$ | $0.0000152^{+0.0000035}_{-0.0000033}$ | $0.0000189^{+0.0000035}_{-0.0000033}$ |
| $R/R_*$ | $0.01847^{+0.00055}_{-0.00056}$ | $0.01842^{+0.00055}_{-0.00053}$ | $0.01833^{+0.00056}_{-0.00057}$ |
| *Kepler-223 d Parameters:* | | | |
| P (d) | $14.78881^{+0.00049}_{-0.00040}$ | $14.78869^{+0.00030}_{-0.00027}$ | $14.78862^{+0.00025}_{-0.00024}$ |
| $T_0$ (BJD-2454900) | $804.8502^{+0.0022}_{-0.0023}$ | $804.8504^{+0.0023}_{-0.0024}$ | $804.8492^{+0.0022}_{-0.0023}$ |
| $e \cdot \cos(\omega)$ | $0.020^{+0.031}_{-0.030}$ | $0.020^{+0.026}_{-0.024}$ | $0.000^{+0.011}_{-0.013}$ |
| $e \cdot \sin(\omega)$ | $0.017^{+0.023}_{-0.076}$ | $0.010^{+0.020}_{-0.032}$ | $-0.001^{+0.015}_{-0.021}$ |
| $|i - 90|$ (°) | $2.02^{+0.29}_{-0.52}$ | $2.06^{+0.26}_{-0.32}$ | $1.68^{+0.30}_{-0.29}$ |
| $\Omega$ (°) | 0.0 (fixed) | 0.0 (fixed) | 0.0 (fixed) |
| $M/M_*$ | $0.0000203^{+0.0000040}_{-0.0000039}$ | $0.0000240^{+0.0000039}_{-0.0000035}$ | $0.0000225^{+0.0000032}_{-0.0000032}$ |
| $R/R_*$ | $0.02791^{+0.00056}_{-0.00064}$ | $0.02800^{+0.00056}_{-0.00059}$ | $0.02756^{+0.00056}_{-0.00058}$ |
| *Kepler-223 e Parameters:* | | | |
| P (d) | $19.72553^{+0.00067}_{-0.00071}$ | $19.72567^{+0.00055}_{-0.00054}$ | $19.72568^{+0.00054}_{-0.00048}$ |
| $T_0$ (BJD-2454900) | $817.5231^{+0.0055}_{-0.0048}$ | $817.5237^{+0.0055}_{-0.0051}$ | $817.5231^{+0.0053}_{-0.0046}$ |
| $e \cdot \cos(\omega)$ | $0.017^{+0.042}_{-0.033}$ | $0.017^{+0.026}_{-0.024}$ | $0.013^{+0.014}_{-0.023}$ |
| $e \cdot \sin(\omega)$ | $0.045^{+0.032}_{-0.077}$ | $0.039^{+0.023}_{-0.032}$ | $0.033^{+0.016}_{-0.023}$ |
| $|i - 90|$ (°) | $1.95^{+0.25}_{-0.45}$ | $2.00^{+0.27}_{-0.27}$ | $1.69^{+0.25}_{-0.24}$ |
| $\Omega$ (°) | 0.0 (fixed) | 0.0 (fixed) | 0.0 (fixed) |
| $M/M_*$ | $0.0000102^{+0.0000044}_{-0.0000042}$ | $0.0000145^{+0.0000039}_{-0.0000036}$ | $0.0000130^{+0.0000031}_{-0.0000029}$ |
| $R/R_*$ | $0.02450^{+0.00076}_{-0.00077}$ | $0.02466^{+0.00074}_{-0.00076}$ | $0.02421^{+0.00069}_{-0.00068}$ |

**Extended Data Table 2**: **Complete Kepler-223 Parameters.** DEMCMC posterior probability estimates and uncertainties for all model parameters at $T_{epoch} = 800.0$ (BJD-2,454,900). Three parameter sets are given with fixed stellar mass: (1) DEMCMC results with eccentricity constraint $\mathcal{C}1$ as described in the text, (2) A subset of the $\mathcal{C}1$ DEMCMC results that only retain solutions stable for $10^6$ years and (3) Laplace angle constraint $\mathcal{C}3$ as described in the text and fixed $\Omega_i = 0$ for $i = b, c, d, e$.

| Planet | Period (d) | $T_0$ (BJD-2454900) | $e$ | $i$ $(°)$ | $\Omega$ $(°)$ | $\omega$ $(°)$ | Mass ($M_{jupiter}$) | Radius ($R_p/R_*$) |
|---|---|---|---|---|---|---|---|---|
| b | 7.384720365879194 | 801.516262774051825 | 0.105758145660053 | 90.701847866139545 | 0.0 | 62.597372675420416 | 0.022730704097050 | 0.015954404145479 |
| c | 9.845453934132928 | 800.146170501596430 | 0.172729064427036 | 90.301811036839879 | 0.0 | 85.015828120049491 | 0.017312231285438 | 0.018346434846992 |
| d | 14.788902636701252 | 804.851045349929109 | 0.037330052890247 | 92.189693102657941 | 0.0 | 76.465729705828863 | 0.019623186719198 | 0.027674878130791 |
| e | 19.726218957815664 | 817.521944355066694 | 0.051464531998599 | 92.056638725826986 | 0.0 | 111.706814565803512 | 0.009576406850388 | 0.024759859857039 |
| *Stellar Parameters:* | 1.125 $M_*$ ($M_\odot$) | | $R_*$($R_\odot$): | 1.744528317200141 | $c_1$: | 0.479330549583184 | $c_2$: 0.2 | *dilute:* 0.11202 |
| b | 7.384583733215798 | 801.513943095097261 | 0.061453702027857 | 91.105539095271382 | 0.0 | 37.604238003695137 | 0.020503806935496 | 0.015793288256059 |
| c | 9.845639757204141 | 800.144691508369419 | 0.112391047984129 | 91.085286013475226 | 0.0 | 86.059011138583742 | 0.019192688432573 | 0.018609959659302 |
| d | 14.788880252356291 | 804.849755312464254 | 0.026604678672708 | 91.966288309512123 | 0.0 | 58.807213313926120 | 0.025560722351934 | 0.028232411829371 |
| e | 19.725687523818440 | 817.519383441790524 | 0.060783217179960 | 91.806556478578258 | 0.0 | 76.156009027159996 | 0.015467248730564 | 0.024265426463497 |
| *Stellar Parameters:* | 1.125 $M_*$ ($M_\odot$) | | $R_*$($R_\odot$): | 1.683974231305496 | $c_1$: | 0.532243950638929 | $c_2$: 0.2 | *dilute:* 0.11202 |

**Extended Data Table 3**: **Best-fit Kepler-223 initial conditions.** Best-fit initial planet conditions found by DEMCMC under $\mathcal{C}1$ (top) and $\mathcal{C}3$ (bottom) constraints at $T_{epoch} = 800.0$ (BJD-2454900) with $\chi^2 = 746480$ and $746489$ respectively.

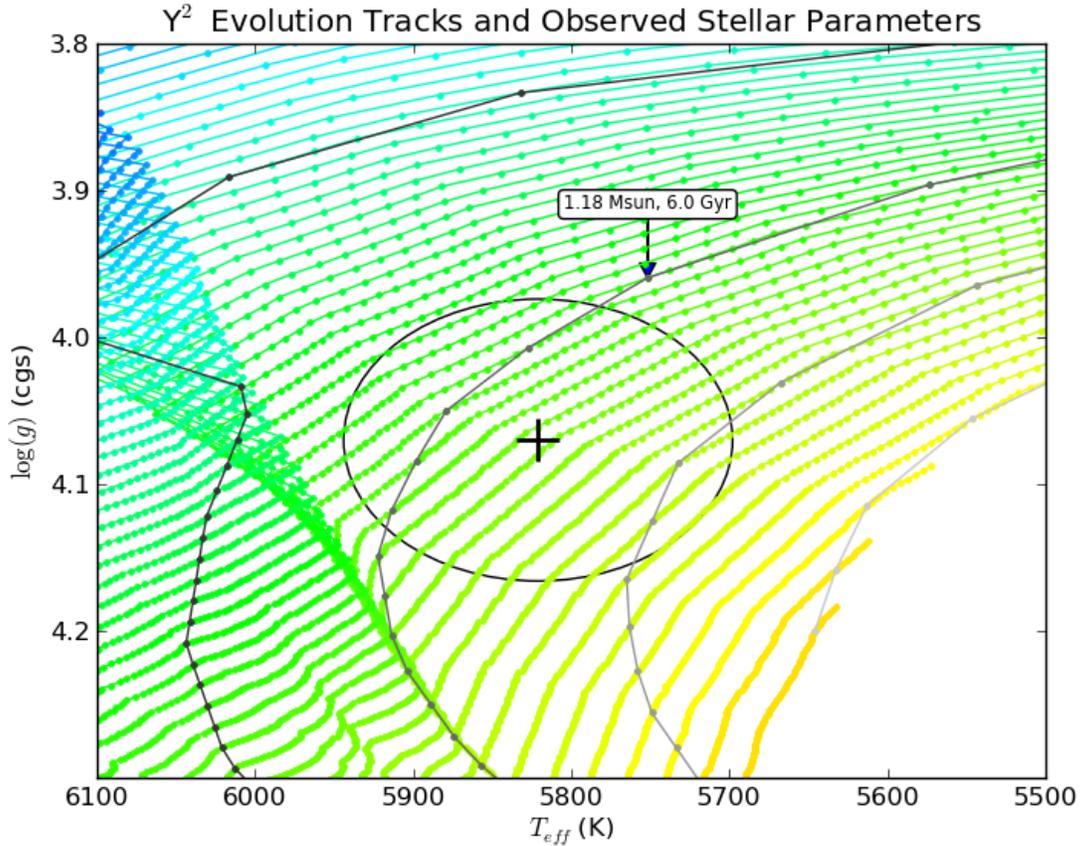

**Extended Data Figure 1**: **Spectroscopic fit of the Kepler-223 star.** A fit to Yonsei–Yale (Y²) evolution tracks (coloured lines) with 0.01-Gyr increments marked with filled circles. Colours correspond to mass with increments of $0.01 M_\odot$ from $1.0 M_\odot$ (orange) to $1.4 M_\odot$ (darkest blue). Isochrones (grey lines) are over-plotted in 2-Gyr increments from 4 Gyr (darkest grey) to 10 Gyr (lightest grey) with filled circles every $0.01 \mathrm{M} M_\odot$ increment. One point is labelled for reference ($M_{\mathrm{Sun}} = M_\odot$). The best-fit ($T_{\mathrm{eff}}$, $\log(g)$) value (black cross) and an ellipse (black) whose semi-major axes indicate $1\sigma$ uncertainties of each parameter found from spectral matching are indicated. The stars in this area of parameter space have evolved off the main sequence.

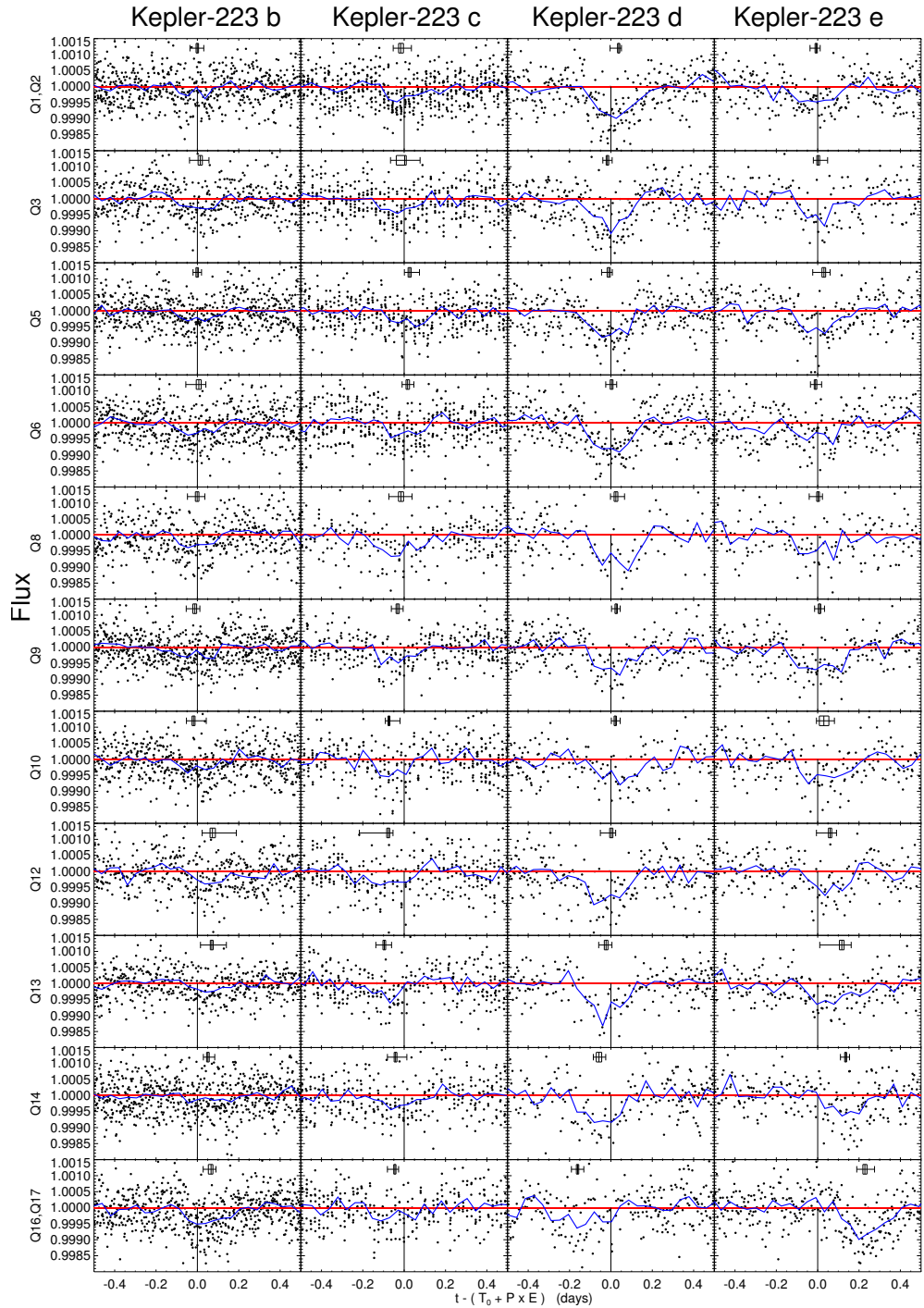

**Extended Data Figure 2**: **Long-cadence light curve for each planet, broken down by quarter (Q).** Data (black filled circles) are binned via a moving average to give the blue curve, to reduce the scatter relative to the horizontal red line indicating no signal. Each panel is centred on the transit times predicted using the linear ephemeris ($T_0$ and $P$) of Batalha et al. (2013)[52] (vertical black lines), with the horizontal axis the time in days from the $E$th predicted transit time. The box-and-whisker error bars indicate the best-fit mid-transit time and $1\sigma$ and $3\sigma$ uncertainties based on $\Delta\chi^2 = 1$ and $\Delta\chi^2 = 9$. $\chi^2$ values are computed by sliding an overall fit to the transit horizontally across the data and interpolating. Their offset relative to the linear ephemeris lines indicates the magnitudes of the TTVs.

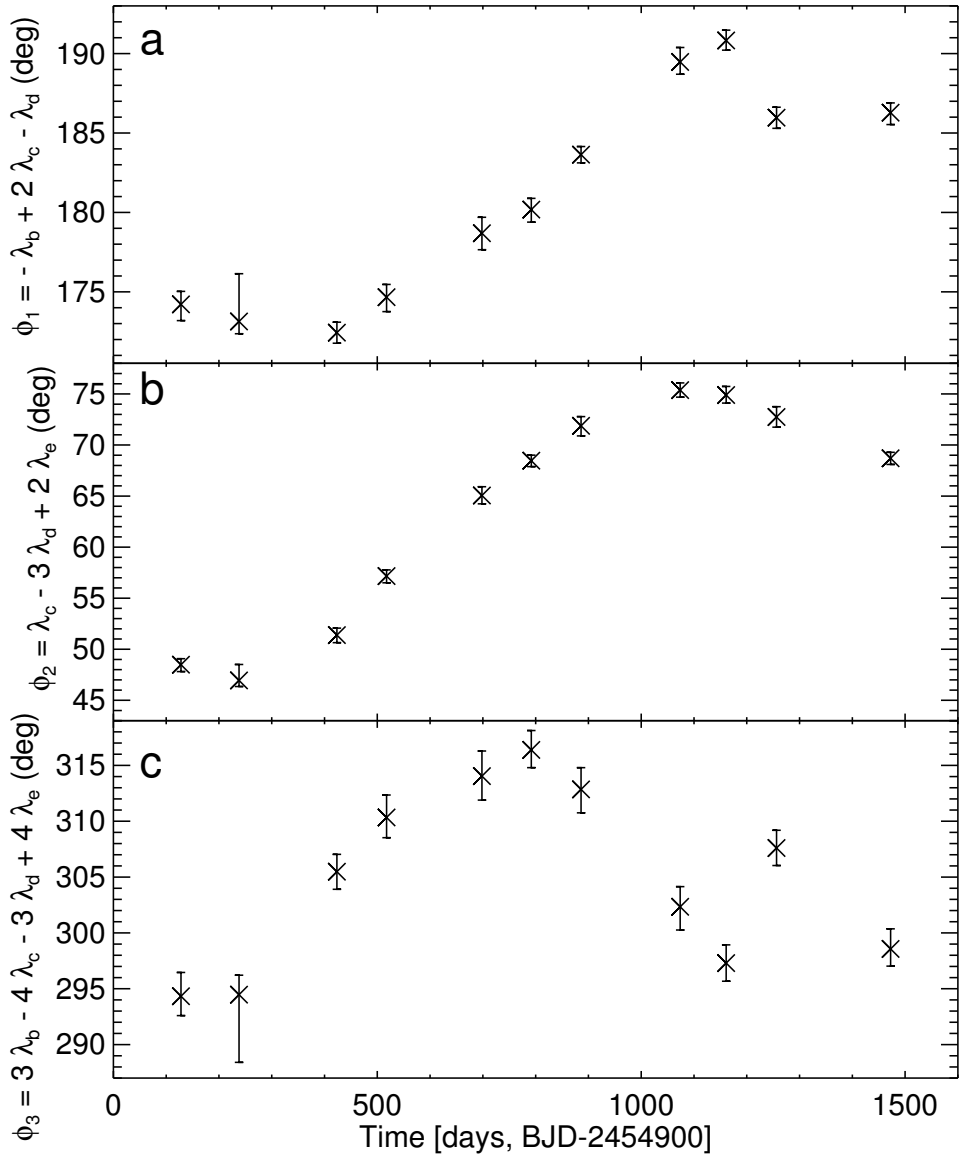

**Extended Data Figure 3**: **Laplace-angle librations detected by binning transits into quarters and assuming zero eccentricity.** Error bars show $1\sigma$ uncertainties based on $\Delta\chi^2 = 1$. Almost a full libration cycle of all angles is observed in the $\sim$1,500-day observing window. The amplitude of oscillation in the four-body Laplace angle ($\phi_3$; **c**) is similar in amplitude to each of the individual Laplace angles ($\phi_1$, **a**; $\phi_2$, **b**). Because $\phi_3 = -3\phi_1 + 2\phi_2$, this amplitude could naively be expected to be much larger; however, $\phi_1$ and $\phi_2$ are closely related, owing to the four-body resonance of the Kepler-223 system, in contrast to two independent three-body resonances.

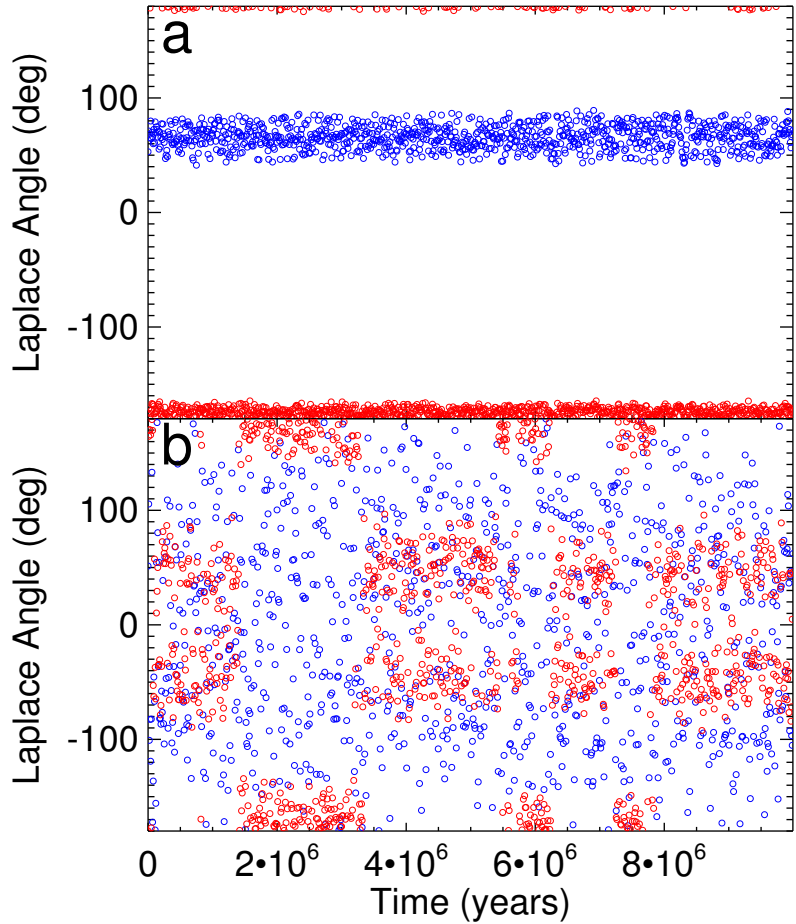

**Extended Data Figure 4**: **Laplace angle variations for two $10^7$ year stable solutions.** **a**, The librating Laplace angles ($\phi_1$, red; $\phi_2$, blue) for a solution from the C3 DEMCMC posterior. Laplace angles librate over the entire $10^7$ years. The orbital-period distribution in Extended Data Fig. **5** uses this model. **b**, Another solution from $\mathcal{C}3$, in which the inner Laplace angle ($\phi_1$; red) librates near the observed value initially, but begins switching chaotically between three different libration centres. This is not uncommon in the $\mathcal{C}3$ DEMCMC posterior. Despite the initial constraint on the outer Laplace angle ($\phi_2$; blue), there are long periods of circulation with intermittent libration.

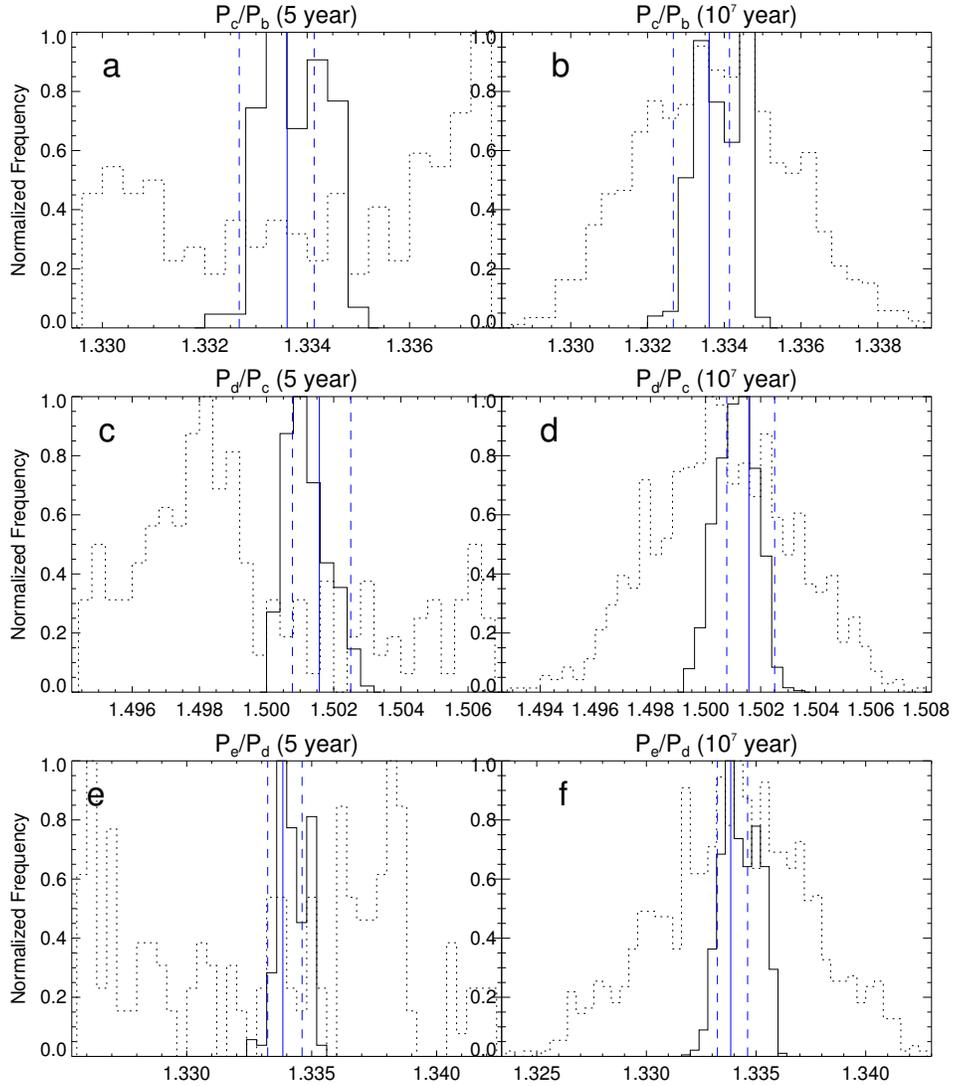

**Extended Data Figure 5**: **Orbital-Period ratios of librating and non-librating solutions fit to data. a,c,e,** The distribution of osculating period ratios for each neighbouring planet pair ($P_c/P_b$, a; $P_d/P_c$, c; $P_e/P_d$, e) over a randomly selected 4-year window in the first $10^4$ years for two $10^7$-year-stable parameter sets from the $\mathcal{C}3$ DEMCMC posterior solution. The dotted histogram represents a solution that showed substantial periods of Laplace-angle circulation. The solid histogram represents a solution in which both $\phi_1$ and $\phi_2$ librate for $10^7$ years. The blue vertical line indicates the empirical mean period; blue dashed vertical lines represent the highest and lowest quarter-to-quarter period measured. **b, d, f,** The same as in a, c, e, but over the entire $10^7$-year interval.

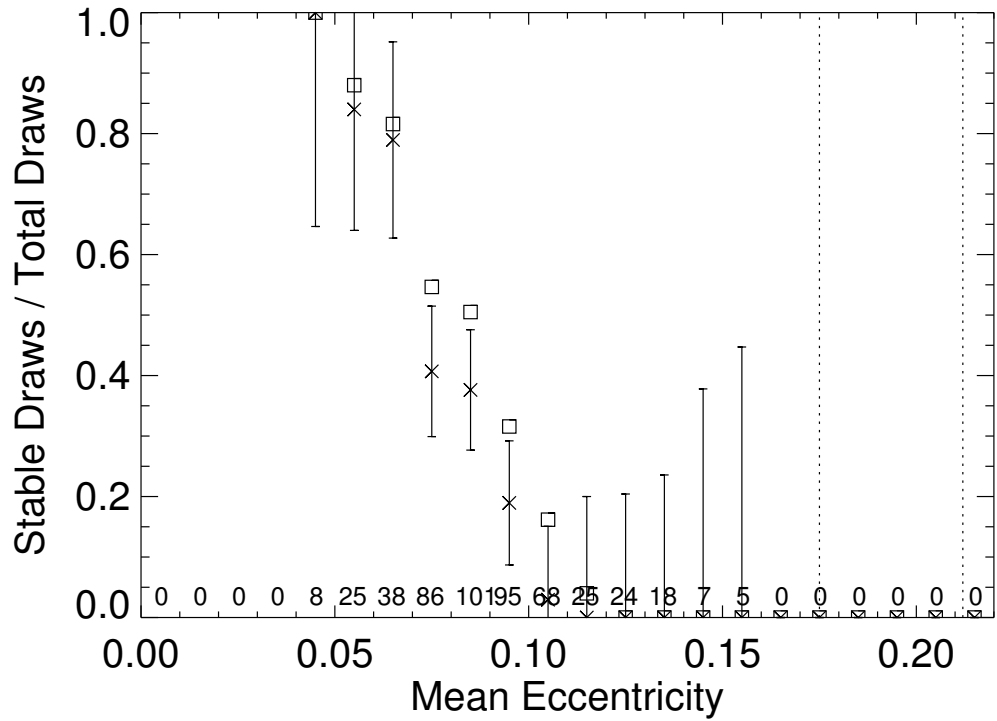

**Extended Data Figure 6**: **System stability as a function of mean planetary eccentricity.** The fraction of 500 random draws from the $\mathcal{C}1$ posterior that survive for $10^7$ years (crosses) and $10^6$ years (squares) as a function of four-planet-mean eccentricity in bins of width 0.01. $1\sigma$ statistical uncertainties are included as vertical error bars on the crosses. Dotted lines indicate the two eccentricity limits for the planets used in $\mathcal{C}1$: 0.175 (planets c and e) and 0.212 (planets b and d). Numbers represent the total number of draws in each eccentricity bin. The fraction of $10^7$-year-stable systems falls sharply and is consistent with zero well below the eccentricity cuts imposed by $\mathcal{C}1$.